\documentclass[sigconf]{acmart}
\usepackage{algorithm}
\usepackage{algorithmic}

\usepackage{amsmath}
\usepackage{listings}
\usepackage[none]{hyphenat}
\usepackage{graphicx}
\usepackage{textcomp}
\usepackage{xcolor}
\usepackage{draftwatermark}

\SetWatermarkText{%
  \textsf{\fontsize{70}{80}\selectfont Accepted by ACM WiSec 2026}
}
\SetWatermarkScale{0.9}
\SetWatermarkColor[gray]{0.83}
\SetWatermarkAngle{45}
\usepackage{tabularx}
\usepackage{array}
\usepackage{url}  
\usepackage{xurl}
\usepackage{microtype}
\usepackage{balance}
\newcommand{\cmark}{\ensuremath{\checkmark}}
\newcommand{\xmark}{\ensuremath{\times}}

\newcolumntype{L}{>{\raggedright\arraybackslash}p{0.17\linewidth}}
\newcolumntype{Y}{>{\raggedright\arraybackslash}X}


\newcommand{\code}[1]{\url{#1}}

\newcommand{\inlinecode}[1]{\lstinline|#1|}

\def\BibTeX{{\rm B\kern-.05em{\sc i\kern-.025em b}\kern-.08em
    T\kern-.1667em\lower.7ex\hbox{E}\kern-.125emX}}

\AtBeginDocument{%
  \providecommand\BibTeX{{%
    Bib\TeX}}}

\copyrightyear{2026}
\acmYear{2026}
\acmConference[WiSec '26]{Proceedings of the 19th ACM Conference on Security and Privacy in Wireless and Mobile Networks}{June 30-July 03, 2026}{Saarbrücken, Germany}
\acmBooktitle{Proceedings of the 19th ACM Conference on Security and Privacy in Wireless and Mobile Networks (WiSec '26), June 30-July 03, 2026, Saarbrücken, Germany}

\begin{document}
\raggedbottom

\title{PrivacyAssist: A User-Centric Agent Framework for Detecting Privacy Inconsistencies in Android Apps}


\author{Tran Thanh Lam Nguyen}
\email{lam.nguyen@uninsubria.it}
\affiliation{%
  \institution{University of Insubria}
  \city{Varese}
   \country{Italy}
}

\author{Edoardo Di Tullio}
\email{editullio@studenti.uninsubria.it}
\affiliation{%
   \institution{University of Insubria}
    \city{Varese}
   \country{Italy}
}

\author{Barbara Carminati}
\email{barbara.carminati@uninsubria.it}
\affiliation{%
  \institution{University of Insubria}
  \city{Varese}
   \country{Italy}
}

\author{Elena Ferrari}
\email{elena.ferrari@uninsubria.it}
\affiliation{%
   \institution{University of Insubria}
  \city{Varese}
   \country{Italy}
}

\renewcommand{\shortauthors}{Lam et al.}

\begin{abstract}

Mobile apps offer significant benefits, but their privacy protections often remain ineffective and confusing for users. While prior work mainly analyzes app privacy vulnerabilities, few approaches help users understand, set, and enforce their privacy preferences. This paper presents PrivacyAssist, a multi-agent LLM-based platform that detects inconsistencies between user-granted permissions and developers’ declared sensitive data collection and sharing practices. Using Retrieval-Augmented Generation (RAG), PrivacyAssist provides concise explanations and real-time on-device warnings to support informed installation decisions. We evaluate PrivacyAssist with 200 users and 2,347 Android apps, finding that only 16\% of apps are fully consistent between granted permissions and declared data practices.
\end{abstract}


\maketitle

\section{Introduction}\label{sec:introduction}

Over the past decade, the mobile application market has grown rapidly, driven by the widespread adoption of smartphones and high-speed internet. Android has become the dominant mobile operating system, accounting for approximately 71.68\% of the global market share.\footnote{\url{https://gs.statcounter.com/os-market-share/mobile/worldwide}}
Its open-source nature through the Android Open Source Project (AOSP) allows developers to freely modify and extend the platform. However, this openness is a double-edged sword, making Android a prime target for privacy attacks compared to closed ecosystems such as Apple’s iOS \cite{karayel2025understanding}.

To protect user privacy, regulations such as the General Data Protection Regulation (GDPR)\footnote{\url{https://gdpr-info.eu/}} and the California Consumer Privacy Act (CCPA)\footnote{\url{https://oag.ca.gov/privacy/ccpa}}
emphasize two core principles: lawfulness, which requires user consent for data collection, and transparency, which requires users to be informed about how and why their data is collected and shared. To support these principles, Android provides two main privacy mechanisms: the \textit{permission model (PM)} and \textit{Data Safety (DS)}. The PM limits app access to sensitive data and system resources, such as location and camera, through permissions declared in the \texttt{AndroidManifest.xml} file and granted by users at runtime (i.e., runtime permission). In contrast, DS requires developers to disclose on Google Play what types of data their apps collect or share and for what purposes. However, both mechanisms remain insufficient because they rely largely on developer-provided declarations and lack verification by Google \cite{nguyen2026alibis}. As a result, apps may request excessive permissions unrelated to their functionality \cite{alkinoon2025comprehensive}, or behave inconsistently with their DS declarations by sending undeclared sensitive data to backend servers or third-party services for storage, analytics, or profiling \cite{nguyen2025detecting}. Moreover, DS disclosures are not automatically shown during app installation, forcing users to actively search for this information on Google Play.

Furthermore, Android permissions have two levels: normal (e.g., \texttt{INTERNET}) and dangerous (e.g., \texttt{ACCESS\_FINE\_LOCATION}).\footnote{Detailed definitions and protection levels of all Android permissions are published in Google’s official documentation: \url{https://developer.android.com/reference/android/Manifest.permission}} 
However, only dangerous-level permissions require user consent through a runtime permission mechanism, while normal-level permissions are automatically granted, and this vulnerability has been exploited by hackers to illegally collect user data \cite{li2020exploiting}.

Additionally, research \cite{prange2024soups} shows that users rarely carefully read privacy-related information and tend to grant permissions to apps based on their preferences, particularly for free features offered by the apps.
One reason for this behavior is that Android permissions are complex and difficult for end users who lack security/privacy knowledge to understand. 
For example, to protect GPS information, Google provides two location permissions: \texttt{ACCESS\_COARSE\_LOCATION} and \texttt{ACCESS\_FINE\_LOCATION} with different precision in gathering location information (700-1000 m vs 3-5 m). 
Although the runtime permission mechanism allows users to agree or refuse permission, it does not clearly explain the practical differences between permission variants.

The above limitations highlight not only technical gaps in Android’s privacy mechanisms but also ambiguities in the underlying threat landscape. 
In this work, we consider privacy threats arising from three main sources: (1) careless or malicious developers who misconfigure permissions in the manifest file and/or request excessive permissions that are not aligned with the app’s functionality; (2) developers who inaccurately disclose the types of sensitive data collected or shared in the DS section; and (3) limited semantic transparency in Android’s permission model.
These challenges collectively motivate the need for mechanisms that can detect inconsistencies between the permissions requested by an app and the sensitive data practices declared in the DS section, while also providing real-time warnings and accessible explanations.

In recent years, Large Language Models (LLMs)~\cite{minaee2024large} have demonstrated strong capabilities for automating complex reasoning tasks, such as compliance analysis and vulnerability assessment. Their ability to generate clear, natural-language explanations makes them well-suited for explaining privacy risks to end users \cite{nguyen2025llms}. Building on these strengths, we propose \textbf{PrivacyAssist}, a user-oriented alert system that runs directly on smartphones to detect inconsistencies between declared privacy practices and runtime-granted permissions.
PrivacyAssist leverages a two-agent architecture: Agent-1, embedded on the device, and Agent-2, deployed on the server. 
Unlike previous approaches that rely on static analysis of permissions declared in the \texttt{AndroidManifest.xml} \cite{nguyen2025detecting}, our system operates in real time and focuses on permissions that are effectively active during execution. Specifically, PrivacyAssist automatically detects newly installed apps and identifies, at runtime, their \textbf{runtime behavior}, that is, the permissions actually granted by the user, including both dangerous-level permissions and normal-level permissions automatically approved by the OS. In parallel, it retrieves the app’s declared information from DS and its public feature description (denoted as \textbf{declared behavior}). PrivacyAssist then compares the runtime behavior with the declared behavior to detect inconsistencies and generates concise, on-device alerts to support user decision-making.

State-of-the-art recent LLM-based approaches primarily focus on policy understanding or app-centric compliance analysis. For example, SEEPRIVACY \cite{pan2024new} generates contextual privacy explanations by analyzing GUI screenshots; however, it depends on user-captured screenshots and cannot handle background behaviors. \cite{cory2026word} improves phrase-level annotation of privacy policies for legal analysis, yet it targets professional compliance assessment rather than end users.
In one of our previous works \cite{nguyen2025detecting}, we combine LLMs and knowledge graphs to analyze network traffic and manifest files to detect whether an app is illegally sharing sensitive data across 14 sensitive data type categories, as defined by Google. 
Additionally, the data's destination (e.g., third-party service or app backend) is examined to detect recipients not declared in the manifest (i.e., deep links). However, the proposed solution is not designed for real-time analysis. Furthermore, it requires rooting the phone, which most users would refuse due to warranty issues, making it suitable only for gathering evidence of violations rather than early warning.
In contrast, PrivacyAssist is specially designed for real-time, on-device deployment. PrivacyAssist's component (i.e., Agent-1) operates directly on resource-constrained smartphones without requiring rooting and focuses on the permissions actually granted by users, thereby providing them early warnings.
However, we inherit the list of 14 sensitive categories along with their corresponding data types from \cite{nguyen2025detecting} to build the external database for RAG (see  Section \ref{sec:implementation}).

The remainder of this paper is organized as follows. Section \ref{sec:background} introduces LLM-based AI agents and related challenges, Section \ref{sec:workflow} presents the PrivacyAssist architecture and workflow, Section \ref{sec:implementation} details the implementation, Section \ref{sec:evaluation} reports the experiments, whereas Section \ref{sec:conclusion} concludes the paper.

\section{AI agents \& their challenges in PrivacyAssist}\label{sec:background}
AI agents have long been studied in the literature across various paradigms \cite{piccialli2025agentai}.
Recently, the emergence of LLMs has led to a new concept, namely, LLM-based AI Agents. 
An LLM-based AI agent is an autonomous system in which LLMs function as the central reasoning component, interpreting contextual inputs, decomposing tasks, generating intermediate decisions, and producing final outputs \cite{kolt2025governing}. 
In PrivacyAssist, we focus specifically on LLM-based AI agents because they are more flexible than traditional AI agents (e.g., reinforcement-learning agents). Particularly, LLM-based AI Agents do not require task-specific retraining from scratch, making them well-suited for user-centric privacy analysis in dynamic mobile environments.
However, LLM-based AI agents still inherit the weaknesses of LLMs, thus creating challenges that need to be addressed. A first limitation concerns the constraints on context length and token budgets, which may prevent the model from jointly considering all relevant information. For example, the input token limit of the Llama-3-8B model\footnote{\url{https://ollama.com/library/llama3}} provided by Meta AI is 8,192 tokens. Additionally, LLMs often provide lengthy responses that can be confusing for users and make it difficult to display all warnings on small smartphone screens.
A second and more critical limitation is represented by the phenomenon of \emph{hallucinations}, i.e., the generation of plausible but unsupported or incorrect statements. In a user-centered context, particularly regarding privacy, such ambiguous behavior may undermine user trust and negatively affect decision-making \cite{tu2020permissionConsistency}.

To address the limitations of token constraints and lengthy responses, PrivacyAssist does not provide raw text information to the LLM. Instead, it provides processed input information in JSON format. Furthermore, PrivacyAssist does not use the entire analysis result of the LLM  to display alerts to the user, as these responses are often lengthy and difficult to follow. Instead, it includes a summarization mechanism that condenses the LLM response into a more concise, understandable form. 
To mitigate hallucinations, we do not rely solely on the LLM to compare runtime behavior with declared behavior and identify the inconsistencies. Indeed, because the LLM relies heavily on training data, when that data becomes outdated, the LLM may make inaccurate inferences, especially given that the Android OS is updated annually. 
Therefore, to enhance the context and inference capabilities of the LLM-based AI agent, we use Retrieval-Augmented Generation (RAG),\cite{sawarkar2024blended} a technique that allows the LLM to supplement information from external knowledge sources.
Thanks to RAG, the LLM not only relies on the training dataset but can also retrieve information related to the input prompt from an external database (ED) to enrich its reasoning. 
In the PrivacyAssist scenario, the input prompt includes information about the app's runtime behavior, i.e., a list of granted permissions. The RAG uses the input prompt to retrieve detailed definitions of each permission from ED via a similarity search. ED is created and updated based on the latest information on Android permissions and DS sections. Then, the search results are combined with the input prompt to enhance the LLM's inference context (see Section \ref{sec:workflow}).

\section{PrivacyAssist Architecture \& Workflow}\label{sec:workflow}
\begin{figure}
   \centering
   \includegraphics[width=\linewidth]{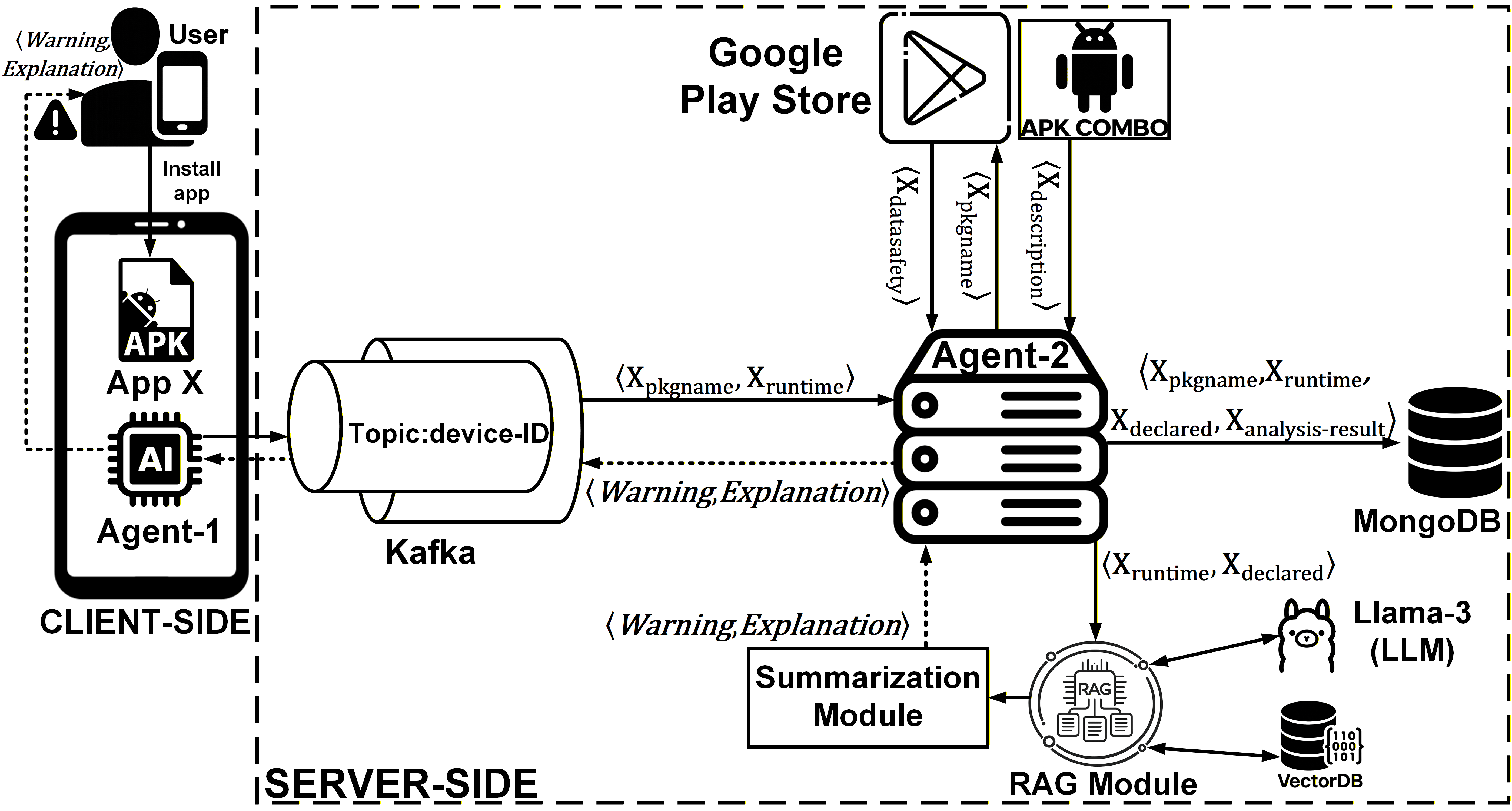}
   \caption{PrivacyAssist architecture}
   \label{figure-architecture}
\end{figure}

This section presents the architecture and workflow of PrivacyAssist along with an illustrative usage example.

\subsection{Architecture}
PrivacyAssist adopts a client–server architecture in which Agent-1 operates directly on the user’s smartphone (client side), while Agent-2 runs on the server side (see Figure \ref{figure-architecture}). 
Specifically, Agent-1 is responsible for collecting app information (i.e., app's package name), device information (i.e., device ID), and runtime behavior (i.e., the list of granted permissions) when the user installs the app. Meanwhile, Agent-2 collects declared behavior, including DS declarations and a general description of the app's features. The process of analyzing and identifying inconsistencies between runtime behavior and declared behavior is performed on the server side by the LLM with enhanced context from the RAG.
Then, LLM's response is concisely and clearly summarized by the summary module and sent back to Agent-1 to display warnings and explanations on the user's smartphone.
We use Kafka,\footnote{\url{https://kafka.apache.org/}} a distributed event-streaming platform designed to handle real-time data flows, to exchange information between Agent-1 and Agent-2. Specifically, Kafka supports multiple logical channels, called topics, for sending and receiving data using a publish/subscribe mechanism.
We leverage this ability to create a separate communication channel for each user, ensuring their data cannot be accessed by others. 
Finally, the app's package name, device ID, runtime/ declared behavior, and analysis results are stored in MongoDB.\footnote{\url{https://www.mongodb.com/}}
This design allows PrivacyAssist to respond quickly without re-analyzing in case the user has previously installed and used the app (e.g., installed, uninstalled, and later reinstalled it).

\begin{table*}
\centering
\caption{PrivacyAssist analysis of Facebook }
\label{tab:facebook_example}
\renewcommand{\arraystretch}{1.12}
\setlength{\tabcolsep}{6pt}
\footnotesize

\newcolumntype{L}{>{\raggedright\arraybackslash}p{0.10\linewidth}}
\newcolumntype{Y}{>{\raggedright\arraybackslash}X}
\begin{tabularx}{\textwidth}{|L|Y|}
\hline
\textbf{Facebook (FB)} & \textbf{Data} \\
\hline

\textbf{$FB_{\text{runtime}}$} &
\code{READ_CALENDAR}, \code{WRITE_CALENDAR}, \code{READ_CALL_LOG}, \code{CAMERA},
\code{READ_CONTACTS}, \code{WRITE_CONTACTS}, \code{READ_EXTERNAL_STORAGE},
\code{ACCESS_FINE_LOCATION}, \code{ACCESS_COARSE_LOCATION}, \code{RECORD_AUDIO},
\code{BLUETOOTH_CONNECT}, \code{BLUETOOTH}, \code{BLUETOOTH_ADMIN},
\code{READ_PHONE_STATE}, \code{READ_BASIC_PHONE_STATE}, \code{READ_PHONE_NUMBERS},
\code{CALL_PHONE}, \code{ANSWER_PHONE_CALLS}, \code{MANAGE_OWN_CALLS}
\\
\hline

\textbf{$FB_{\text{datasafety}}$} &
Files and docs; App activity; Audio; Photos and videos; Health and fitness; Personal info;
Web browsing; App info and performance; Calendar; Financial info; Contacts; Messages;
Device or other IDs; Location
\\
\hline

\textbf{$FB_{\text{description}}$} &
Facebook is a social networking app for discovering content (personalized Feed, Reels, search,
Marketplace, Meta AI), joining communities, messaging friends, making voice/video calls, and
creating/sharing posts, stories, and stickers/images.
\\
\hline

Analysis result &
\textbf{Matched mappings:}
\code{READ_CALENDAR}, \code{WRITE_CALENDAR} (Calendar~\cmark);
\code{READ_CONTACTS}, \code{WRITE_CONTACTS} (Contacts~\cmark);
\code{ACCESS_FINE_LOCATION}, \code{ACCESS_COARSE_LOCATION} (Location~\cmark);
\code{RECORD_AUDIO} (Audio~\cmark);
\code{CAMERA} (Photos and videos~\cmark);
\code{READ_EXTERNAL_STORAGE} (Photos and videos / Files and docs~\cmark);
\code{READ_PHONE_STATE}, \code{READ_BASIC_PHONE_STATE}, \code{READ_PHONE_NUMBERS}, \code{BLUETOOTH_CONNECT}, \code{BLUETOOTH}, \code{BLUETOOTH_ADMIN} (Device or other IDs~\cmark)\par\smallskip

\textbf{Detected inconsistencies:}\par
\textbf{Case 1}: \code{READ_CALL_LOG}, \code{CALL_PHONE}, \code{ANSWER_PHONE_CALLS}, \code{MANAGE_OWN_CALLS}
$\leftrightarrow$ (no declared Data Safety category \textit{Call logs})~\xmark\par 
\textbf{Case 2}: Web browsing; App activity; App info and performance; Messages; Personal info;
Financial info; Health and fitness $\leftrightarrow$
(collected but no corresponding granted runtime permissions)~\xmark\par
\textbf{Excessive data collection:} The app belongs to the \textit{Social} category, and its core features do not relate to collecting \textit{Web browsing} and \textit{Health and fitness}, which may increase the risk of over-collection.
\\
\hline
Warning \& explanation &
The app can collect your call history, but the app developer does not mention it in the privacy information shown on Google Play.\par
The app states that it collects data such as your browsing activity, app usage details, messages, personal information, financial information, and health data, even though it has not received your consent to access this information on your device.\par
The app may over-collect because web browsing, financial, and health data may not relate to its Social features.
\\
\hline

\end{tabularx}
\end{table*}

\subsection{Workflow}

Let $X$ denote an app installed by user $U$  from the app store (e.g., Google Play Store).
To generate privacy warnings, PrivacyAssist operates in a four-stage workflow, including (1) initialization, (2) monitoring, (3) analysis, and (4) warning and explanation, which we explain below.

\textbf{Initialization stage.}
After Agent-1 is installed and run on user $U$'s smartphone, it determines the device ID  (denoted as $U_\text{deviceID}$) and uses it to create a Kafka topic that serves as the communication channel for exchanging data with  Agent-2.

\textbf{Monitoring stage.}
Agent-1 automatically captures the installation event when the user installs app $X$ from the app store and determines the package name of app $X$ (denoted as $X_\text{pkgname}$), that is, its unique identifier. 
Next, Agent-1 determines the runtime behavior of app $X$ (denoted as $X_\text{runtime}$) by querying the Android PackageManager API to retrieve the set of permissions effectively granted to $X$ during execution, including both user-approved dangerous permissions and automatically granted normal permissions (see Section \ref{sec:implementation}).
Finally, Agent-1 sends $\langle X_{\text{pkgname}}, X_{\text{runtime}} \rangle$ to Agent-2 via the Kafka topic created in the initialization stage.

\textbf{Analysis stage.}
After receiving $X_\text{pkgname}$ and $X_\text{runtime}$  from Agent-1, Agent-2 uses $X_\text{pkgname}$ to retrieve: 1) from Google Play Store, the list of sensitive data categories that $X$  collects and shares, as declared in the DS (denoted as $X_\text{datasafety}$), and 2) a general description of $X$ (e.g., Gmail provides email services and belongs to the \texttt{Communication} category) from APKcombo.\footnote{Since Google Play provides DS details but not comprehensive app descriptions or categories, we rely on APKCombo as an alternative source, as it is a widely accessible and reputable global app store \url{https://apkcombo.com}} All this information forms the declared behavior of app $X$ (denoted as $X_\text{declared}$).
Next, we identify 3 cases that lead to inconsistencies between runtime and declared behavior and we use them with $X_\text{runtime}$ and $X_\text{declared}$ to build an input prompt for RAG as follows: 

$P$ = \textit{``\textbf{Task:} Detect inconsistencies between runtime-granted permissions and declared DS information according to the following definitions.\\
\textbf{Inputs:} $X_\text{runtime}$ (permissions granted to app $X$) and $X_\text{declared}$ (sensitive data categories the app claims to collect).\\
\textbf{Inconsistency cases:} Case~1 occurs when permission to access sensitive data is granted in $X_\text{runtime}$  but not declared in $X_\text{declared}$; Case~2 occurs when sensitive data is declared in $X_\text{declared}$ without corresponding granted permissions in $X_\text{runtime}$; and Case~3 occurs when both Case~1 and Case~2 are satisfied".}

The RAG follows a three-stage process: \textit{preparation}, \textit{retrieval}, and \textit{generation}. 
In the preparation stage, we construct an external database ($ED$) that contains both (1) detailed definitions of all Android permissions officially provided by Google and (2) 14 categories of sensitive data and their corresponding data types \cite{nguyen2025detecting}. This external data source is transformed into embedding vectors $V_{ED} = \text{Embed}(ED)$ and stored in a VectorDB.
In the retrieval stage, the input prompt $P$ is converted to an embedding vector $V_P = \text{Embed}(P)$. Then, $V_P$ is used to perform a similarity search over  VectorDB, retrieving the most relevant vectors $\hat{V}_P$.
In the generation stage, $\hat{V}_P$ is decoded to text form (i.e., $D(\hat{V}_P)$). Finally, we combine $D(\hat{V}_P)$ and input prompt ($P$) to form $\text{Final}_{\text{prompt}} = P + D(\hat{V}_P)$, which is then fed into the LLM to semantically compare $X_\text{runtime}$ and $X_\text{declared}$ and identify potential inconsistencies, if any.
Finally, $X_\text{pkgname}$, $X_\text{runtime}$, $X_\text{declared}$, and the analysis results ($X_\text{analysis-result}$) are stored in MongoDB for quick retrieval without re-analysis in case $X$ is uninstalled and later reinstalled.

\textbf{Warning and explanation stage.}
$X_\text{analysis-result}$ is processed by the summarization module, including 2 tasks: (1) translates technical findings related to inconsistencies between granted permissions and declared DS into non-technical language, and (2) explains potential risks of over data collection w.r.t. the app’s category and features. Then, final user-oriented warnings and explanations are sent back to Agent-1 via Kafka and displayed as an on-device pop-up notification.

\subsection{Illustrative example}
In this section, we use Facebook (FB), a popular app with over 10 billion downloads, to illustrate how PrivacyAssist operates.

\textbf{Initialization stage.}
Assume that the smartphone’s device ID is the string \texttt{9774d56d682e549c}. Agent-1 uses this value to create a Kafka topic for communication between Agent-1 and Agent-2 named \texttt{device-9774d56d682e549c}.

\textbf{Monitoring stage.}
When the user installs FB, Agent-1 identifies $FB_\text{pkgname}$ as \textit{com.facebook.katana}. Let's assume that user $U$ is unconcerned about privacy and grants all permissions requested by FB upon installation. Agent-1 fills $FB_\text{runtime}$ with the permissions shown in row 1 of Table \ref{tab:facebook_example}. Then, it sends $\langle FB_{\text{pkgname}},\, FB_{\text{runtime}} \rangle$ to Agent-2 via Kafka topic \texttt{device-9774d56d682e549c}.

\textbf{Analysis stage.}
Agent-2 uses $FB_\text{pkgname}$ to retrieve $FB_\text{datasafety}$ from the Google Play Store (row 2 of Table \ref{tab:facebook_example}) and $FB_\text{description}$ from APKCombo (row 3 of Table \ref{tab:facebook_example}) to form $FB_\text{declared}$.
Next, Agent-2 inputs $FB_\text{runtime}$ and $FB_\text{declared}$ into the input prompt of the RAG workflow. Specifically, RAG queries $ED$ to obtain detailed definitions of each permission in $FB_\text{runtime}$ and the types of sensitive data that FB claims to collect and share, as stated in its DS section.
Finally, the LLM compares $FB_\text{runtime}$ and $FB_\text{declared}$, leveraging the additional context provided by RAG, and gets the analysis result (Row 4 of Table \ref{tab:facebook_example}).
In the analysis result, the \textit{``matched mappings"} section lists the runtime permissions that align with the declaration in the DS.
In contrast, \textit{``detected inconsistencies"} section indicates that FB falls under Case 3, meaning that it both requests permissions to access certain sensitive data that are not disclosed in the DS  (i.e., call logs), as well as it declares the collection of specific sensitive data categories without requesting the corresponding permissions (i.e., Web browsing, App activity, App info \& performance, Messages, Personal info, Financial info, and Health \& fitness). Finally, in the \textit{``Excessive data collection"} section, LLM identifies the risk of excessive data collection unrelated to FB's features.

\textbf{Warning and explanation stage.}
After summarizing the analysis results, the warnings and explanations are sent to Agent-1 via the Kafka topic and displayed as a pop-up on the smartphone. (row 5 of Table \ref{tab:facebook_example})

\section{Implementation}\label{sec:implementation}
In this section, we present how we built the external database ($ED$) for the RAG workflow and the components of PrivacyAssist.

\subsection{Building external database ($ED$) for RAG}
We use Selenium,\footnote{\url{https://selenium-python.readthedocs.io/}} a framework that automatically simulates human behavior when interacting with a web browser, to crawl permission-related data from Google's official website.\footnote{\url{https://developer.android.com/reference/android/Manifest.permission}}
We perform this step for the 5 most recent Android versions (10-15), then filter out duplicate information to obtain a final list of Android permissions. Specifically, for each permission, we collect its name, its description, and whether it is a dangerous-level or normal-level permission.
Next, we rely on the approach presented in \cite{nguyen2025detecting} to identify Google's definition of 14 categories of sensitive data and their corresponding data types, using again  Selenium to extract this information from Google's official documentation. For example, for the Financial info category, we extract the following info:
\texttt{Financial info = <User payment info: Details of financial accounts (e.g., credit card numbers); Purchase history: Records of user transactions; Credit score: Information about a user's credit rating.>}.
Finally, all collected information is transformed into embedding vectors using an embedding model and then stored in a vector database (i.e., FAISS).\footnote{\url{https://ai.meta.com/tools/faiss/}}

\subsection{PrivacyAssist components}
Agent-1 is implemented in Kotlin for Android, while Agent-2, including the RAG and summarization modules, is implemented in Python.

Agent-1 operates as a background service by exploiting the Android OS's Foreground Service mechanism, ensuring continuous monitoring (see Section \ref{sec:workflow}). Furthermore, to reduce interruptions, Agent-1 is developed with a restart strategy (\texttt{START\_STICKY}), allowing auto-recovery in case of forced stops (e.g., shutdown or OS restart). 
App installation events from the app store are tracked and captured via a Broadcast Receiver, which allows Agent-1 to identify the app's package name. 
Next, Agent-1 uses the Android PackageManager API to query the system permission state associated with the target app and retrieve the list of requested permissions and their grant status (i.e., granted or denied), considering only those actually granted to the app.
This ensures the analysis reflects actual runtime consent rather than static manifest declarations.

Agent-2 uses the Selenium framework to collect the app’s DS information from the Google Play Store and its general functional description from APKCombo (i.e., declared behavior).
Next, we use Langchain\footnote{\url{https://docs.langchain.com/oss/python/langchain/overview}} to build the RAG workflow. LangChain is an open-source framework for developing LLM-based applications, providing built-in functions for flexible connection of LLMs with external data. Furthermore, LangChain supports various LLMs, both open-source (e.g., Llama3) and closed-source (e.g., OpenAI's GPT-4).
For PrivacyAssist, we use the Llama-3-8B  model to ensure that sensitive user data (i.e., the list of apps installed on a smartphone) is not sent to third parties. Additionally, this design allows PrivacyAssist to be an open source solution.\footnote{Source code is published at \url{https://github.com/research-mobile-security/PrivacyAssist.git}}
Finally, we use LangChain again to implement the summary module. Specifically, we use LangChain's MapReduce mechanism to summarize the analysis results (e.g., row 4 in Table \ref{tab:facebook_example}). In the map step, the summarization module decomposes the output of the analysis into components such as \textit{``matched mappings”}, \textit{``detected inconsistencies”}, and \textit{``excessive data collection”} and generates intermediate summaries for each component. Then, during the reduce step, these intermediate results are merged into the final summarization, that is, a brief \textit{``warning \& explanation"} for the user (e.g., row 5, Table \ref{tab:facebook_example}). 
\section{Experimental Evaluation}\label{sec:evaluation}

To evaluate user experience with PrivacyAssist, we recruited 200 participants on a voluntary basis, including undergraduate students and independent volunteers from Italy and Vietnam.
The majority of users were young and middle-aged adults, specifically (1) 18–24 years old (30.8\%), (2) 45–54 years old (23.1\%), (3) 25–34 years old (19.2\%); (4) $\geq$55 years old (15.4\%), and 35–44 years old (11.5\%). 
We prepared a survey questionnaire focusing on two main aspects.\footnote{The questionnaire is available at: \url{https://github.com/research-mobile-security/PrivacyAssist/blob/main/survey.pdf}}
The first part assesses participants’ background, including their knowledge of security and privacy and their app usage habits (e.g., the frequency of checking the app's DS section before installing it).  The survey results show that most participants have basic (46.2\%) or intermediate (38.5\%) security knowledge, while only 15.4\% rate themselves as advanced. 
The second part evaluates users' experiences after using PrivacyAssist, including the clarity of warnings and explanations, the impact on the installation decision, and overall satisfaction with the tool.

Users are instructed to install Agent-1 on their Android smartphone. Then they are free to choose 15 apps from the Google Play Store that they want to use. For participants without an Android device, we provide a testing environment via an Android Virtual Device (AVD).
Next, participants install the selected apps and grant permissions according to their usual usage habits. PrivacyAssist analyzes any inconsistencies between the app's runtime and its declared behavior, then issues warnings and explanations to users, as outlined in Section \ref{sec:workflow}.
Finally, each participant is asked to complete the survey to conclude the experiment.

With 200 test participants, we have a dataset of 3,000 apps. After removing duplicate apps, the \textit{evaluation dataset} has 2,347 apps. In the evaluation dataset, the most frequently selected app categories were: Social Media (21.4\%), Productivity (18.7\%), Games (16.9\%), Communication (14.2\%), and Shopping (9.8\%), while the remaining apps are distributed across the fields of Finance, Health \& Fitness, Education, and Travel. Among the apps in the dataset, 62.5\% of apps have more than 100,000 downloads, and 28.3\% exceed 1 million downloads, indicating that the dataset primarily consists of widely used, popular apps.
This dataset is sufficient to demonstrate the feasibility of our approach, as PrivacyAssist is scalable and not constrained by the number of evaluated apps. As more apps are analyzed, the external database ($ED$) used for RAG is continuously enriched, improving contextual understanding and detection performance.
The  results of the  analysis  show that only 376 apps (16.02\%) fully match between the granted permissions and the DS declarations, while the remaining 1,971 apps (83.98\%) show inconsistencies.
To remind, inconsistencies arise in 3 cases: Case 1: access to sensitive data is granted but not declared; Case 2: sensitive data is declared without corresponding permissions; and Case 3: both above cases occur simultaneously.
Among the 1,971 apps with inconsistencies, 751 (38.11\%) belong to Case 1, 563 (28.56\%) belong to Case 2, and 657 (33.33\%) belong to Case 3.
The Social category has the majority of inconsistencies (95.8\%), followed by Games (88.8\%) and Entertainment (87.8\%), whereas the Education category has the fewest inconsistencies (58.4\%).
The most frequently omitted sensitive data categories in the DS declarations of the evaluation dataset are: Personal (37.42\%), Financial info (27.23\%), Location (21.86\%), and Photos \& Videos (13.49\%).
These findings indicate that inconsistencies between granted permissions and DS are widespread, even in popular apps with high installation rates.

Next, we evaluate the accuracy of PrivacyAssist. We select 100 apps from the evaluation dataset and manually analyze their inconsistencies and excessive data collection unrelated to the app's functionality. The testing team consists of four master students with Android programming experience (25 apps each). We compare the manual analysis results with those obtained by our LLM-based agent.
Results show that for 90\% of the apps, the LLM returns accurate results. For 10\% of the apps, the LLM identifies some false inconsistencies and excessive data collection. 
This mainly occurs in apps with multiple functionalities spanning different categories (e.g., LinkedIn can be classified as a social network, community platform, and business tool). Due to limited contextual understanding of such multi-role apps, the LLM may incorrectly flag certain data types (e.g., files, photos, or videos) as excessive, even though they are functionally relevant. 
However, we do not find any case where the LLM misses manually identified inconsistencies or excessive data collection practices, which would represent the most critical error in our scenario.

In addition, to better understand the impact of the RAG component, we perform an ablation study. Without RAG, PrivacyAssist exhibits a higher rate of false positives (about 35\%) due to the lack of contextual grounding about app functionalities, permission definition, classification of sensitive data types, and data usage practices. 
These findings highlight the importance of RAG in providing domain-specific context.

We then evaluate the experience  users have by using PrivacyAssist. It is important to note that while 44.2\% of participants say they are concerned about sensitive data collection, their behavior before testing PrivacyAssist contradicts this. Only 19.2\% regularly check app permissions, and just 9.6\% consistently review the DS, whereas nearly 60\% rarely or never look at it. In addition, 42.3\% find DS information unclear, 38.5\% have no strong opinion, and only 13.5\% consider it clear. These results highlight careless user behavior and the ineffectiveness of current transparency mechanisms.
In contrast, PrivacyAssist receives strong user acceptance. All participants find it useful (59.6\% very useful; 40.4\% helpful), and report that the warnings and explanations are easy to understand. Importantly, 73.1\% say they would cancel an app installation after receiving a warning.

Finally, we measure the runtime overhead of Agent-1 using Android Studio Profiler\footnote{\url{https://developer.android.com/studio/profile}} under real-world conditions. Since Agent-1 operates continuously in the background, CPU and battery usage are critical. Over 24 hours, Agent-1 consumes on average 1.8\% CPU and about 10\% battery.
Agent-2, running on the server side, takes about 1.5 minutes on average to generate a warning and explanation.\footnote{The 1.5-minute latency is measured from app installation to the display of the warning during the first run, when PrivacyAssist does not have the app’s DS. In subsequent installations, once the DS data is available in the database, warnings are generated more quickly, in only 15 seconds.}
Overall, these results show that PrivacyAssist has minimal overhead on the smartphone, making it practical for real-world deployment.

\section{Conclusion and future work}\label{sec:conclusion}
In this paper, we propose PrivacyAssist, a novel user-centric AI agent platform designed to enhance privacy awareness in the Android ecosystem. 
PrivacyAssist helps users make informed decisions by detecting mismatches between granted permissions and declared data collection practices. 
As future work, we plan to extend PrivacyAssist by embedding an LLM directly into an Android device, thereby eliminating the need to transmit potentially sensitive data off-device and strengthening user privacy guarantees. 
In addition, we plan to further improve our evaluation methodology by incorporating cross-validation, where a subset of apps will be independently assessed by multiple evaluators to measure the level of agreement among evaluators.
Finally, we aim to enhance the current runtime permission mechanism with a dynamic approach that adapts to individual user preferences.


\bibliographystyle{ACM-Reference-Format}
\bibliography{references}

\end{document}